\documentclass[aps,superscriptaddress,twocolumn,showpacs,floatfix,notitlepage,longbibliography]{revtex4-1}
\usepackage{amsmath}
\usepackage{amsfonts}
\usepackage{amssymb}
\usepackage{hyperref}
\usepackage{graphicx}
\usepackage{color}
\usepackage{xcolor}
\usepackage{siunitx}
\usepackage[sort&compress]{natbib}

\begin{document} 
\title{Detuning control of Rabi vortex oscillations in light matter coupling} 
\author{Amir Rahmani}
\affiliation{%
	Department of Physics, Yazd\ university, Yazd, Iran}%

%\date{\today}

\begin{abstract}
  We study analytically the dynamics of vortices in strongly coupled
  exciton--photon fields in the presence of energy detuning. We derive
  equations for the vortex core velocity and mass, where they mainly
  depend on Rabi coupling and the relative distance between the vortex
  cores in photon and exciton fields, and as the result core positions
  oscillate in each field. We use Magnus force balanced with a Rabi
  induced force to show that the core of the vortex behaves as an
  inertial-like particle. Our analysis reveals that the core is
  lighter at periphery of the beam and therefore it is faster at that
  region. While detuning induces oscillations in population imbalance
  of components through relative phase between coupled fields, in
  the presence of topological charges detuning can control the orbital
  dynamics of the cores. Namely, it brings the vortex core to move on
  larger or smaller orbits with different velocities, and changes
  angular momentum and energy content of vortex field.
\end{abstract}

\maketitle
\section{introduction}

The rotation of objects has always been an extremely fascinating and
challenging problem. Examples of angular motions are given in
classical physics \cite{Kleppner13,Saffman92} and engineering
\cite{Meriam12}. When it comes to rotation in a fluid \cite{Childs11},
the term ``vortex'' is used to describe the peculiar motion around a
central region. Vortices typically occur in the wake of fluid motion
where viscosity is the underlying physical mechanism. To quantify the
rotation of a fluid, one can introduce the concept of vorticity,
defined as the curl of the velocity field, and circulation, defined as
the contour integral of the tangential velocity component. The latter
concept is connected to the former (total vorticity) by Stokes's
theorem.  In contrast to a conventional fluid that can have arbitrary
circulation, a Bose--Einstein condensate (BEC) is constrained to carry
quantized circulation. Such a property is on the other hand typical of
a complex--valued wavefunction, where the density profile can include
a depleted region surrounded by quantized circulations. The flow in a
condensate has the intrinsic property of being inviscid, i.e., there
is no viscosity to bring the flow to a stop. Mathematically, a
condensate with a vortex is not simply-connected and then Stokes'
theorem does not apply. The idea of quantized vortices was introduced
for the first time with superfluid Helium, and was then extended to
other condensates. Quantized vortices are now commonplace in
superconductors \cite{blatter_vortices_1994}, superfluids
\cite{leggett_superfluidity_1999}, atomic BECs \cite{Matthews1999}, and polariton
condensates \cite{Lagoudakis2008,domicini18a}. Quantized vortices are
recognized by a null density region around which the phase wraps
itself in quantized units of $2\pi$. Fields with such phase defect
carry a quantized orbital angular momentum \cite{yao2011}. In this
work, we study vortices in two-component (exciton-photon) condensates,
which leads to an interesting interplay for the topology of the fields
due to strong coupling (Rabi oscillations) on the one hand and the
vorticity on the other hand. Specifically, we will focus on the role
of the energy detuning between the exciton and photon components,
which is known to result in nontrivial
dynamics~\cite{Voronova15,Rahmani16}. For such an investigation,
microcavity polaritons~\cite{Kavikin16}---quasiparticles that arise
from the strong coupling between microcavity photon and exciton
quantum well---are ideally suited (one could also consider spin-orbit
coupled BEC~\cite{khamehchi17a,colas18a}). Polaritons have some
interesting attributes. They have a very light effective mass (due to
their photonic component), they interact through their excitonic
component, and their photon and exciton components can also be easily
detuned. The condensed phase of polaritons was reported in 2006
\cite{Kasprzak2006} which has initiated an intense followup of
fundamental research, including quantum hydrodynamics
\cite{Amo2011,Pigeon2011}, superfluidity
\cite{Amo2009,Lagoudakis2008}, Josephson effects \cite{Aba13}, and
vortices
\cite{Lagoudakis2008,Lagoudakis2009,Sanvitto2010,dominici_vortex_2015,Dominici18-vrtex},
among others. The regime of strong coupling, typically evidenced
through the splitting between an upper and a lower polariton branch in
the dispersion relation, is more notable for our purpose for the
dynamics of Rabi oscillations between the photonic and excitonic
components, whose coherent control is now well implemented at the femtosecond level
with polaritons \cite{Dominici2014}.
\begin{figure}[b]
	\begin{center}
		\includegraphics[width=\linewidth]{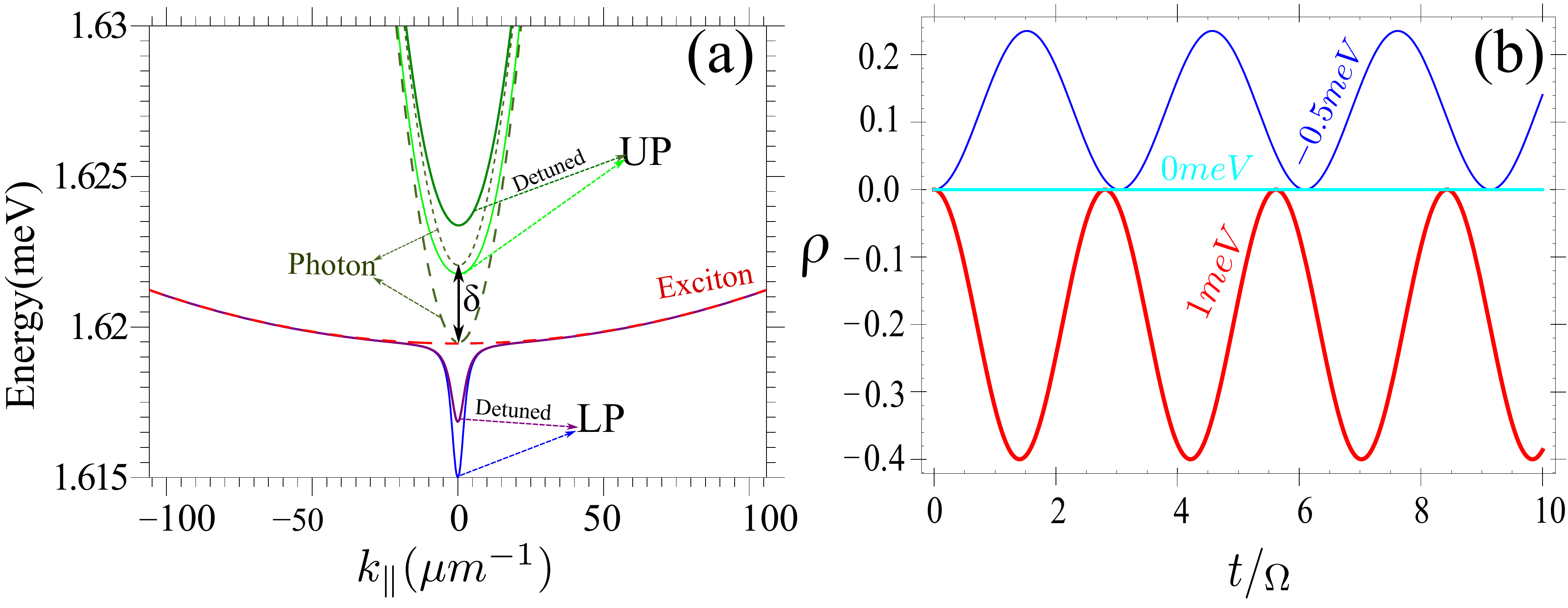}
		\caption{(a): Polariton dispersion for positive
                  detuning, with an upward shift of the branches. (b):
                  Effect of detuning on the population imbalance
                  $\rho=(N_{ph}-N_{ex})/(N_{ph}+N_{ex})$, for three
                  detuning energies: $\delta=-0.5\mathrm{meV}$ in red,
                  $\delta=0\mathrm{meV}$ in cyan, and
                  $\delta=1\mathrm{meV}$ in blue color. $\Omega$ is the
                  Rabi frequency.}
		\label{fig:rytuwteitrgywey89382}
	\end{center}
\end{figure}  

Even fields with no phase defects (vortices) exhibit peculiar
dynamical effects in presence of an energy detuning between the components,
such as oscillations in their relative phase that result in
oscillations in their population imbalance
\cite{Voronova15,Rahmani16}. A particular example is shown in
Fig.~\ref{fig:rytuwteitrgywey89382}. Here, the population imbalance
between the components is given by
$\rho\equiv(N_{ph}-N_{ex})/(N_{ph}+N_{ex})$, with $N_{ph}$ and $N_{ex}$ the
population for the photon and exciton fields respectively, being
initially zero: $N_{ph}(0)=N_{ex}(0)$. For zero energy detuning, there
is no oscillations in $\rho$, although with the imbalance in the initial
relative phase of the components there would be oscillation in
populations. However, by increasing or decreasing the
detuning, some oscillations are triggered. With detuning, the
population imbalance takes a nonzero mean value, which increases with
detuning; it also changes the frequency of oscillation.

In this work, we study analytically how the vortex dynamics gets
affected by detuning in the regime of strong coupling regime. We
restrict our analysis to an ideal polariton gas in the Hamiltonian
regime, where the Rabi coupling is the predominant interaction that
drives the vortex core to move periodically. To describe the core
motion, we invoke a Magnus--like force \cite{Thouless96,Ao93} that is
balanced by a force which originates from the Rabi coupling. This
provides an expression for the inertial  mass of the vortex core, which
changes in time and is of the order of $\SI{e-15}{\kilo\gram}$ and
$\SI{e-12}{\kilo\gram}$, depending on the position of the core. Such a
behavior is particular to a polaritonic system, as its binary nature
gives the core a time-dependent attribute. Based on our analysis, we
demonstrate that the core behaves as an inertial particle, that is, it
has a position, a velocity, and a mass. Furthermore, we are dealing with a not--constant mass case. We show that by changing the
detuning, the core moves faster or slower, which is the result of a
particle transfer between the two coupled fields; then detuning
addresses a controlling mechanism on the dynamics of the core. Also,
it controls the angular momentum of the field.

The text is organized as follows: Section \ref{se:gguger3492hwuwe}
presents the relevant theoretical description of polariton BEC based
on its photonic and excitonic components. In section
\ref{se:jhgyhsyt84347udgi}, we describe the oscillation of the core
and how it is affected by detuning. In the following section
\ref{se:juyftg73493jew92302}, we focus on the effect of detuning on
quantum averages. Finally we present concluding remarks in section
\ref{sec:juhy76tdfdr43gshwqw00}.

\section{Theory}\label{se:gguger3492hwuwe}

The polariton in the exciton--photon basis is an example of a
two--component Bose system. The equations of motions for fields in
binary Bose system can be conveniently described as quantized fields
\cite{Mahan2000} with the following Hamiltonian:
\begin{align}\label{eq:jhyh877hehyi8o3}
\hat{H}&=\frac{\hbar^2}{2m_{ph}}|\vec\nabla\psi|^2+E_{ph}\psi^\dagger\psi+\frac{\hbar^2}{2m_{ex}}|\vec\nabla\varphi|^2+E_{ex}\varphi^\dagger\varphi\nonumber\\&+\hbar \Omega\big( \psi^\dagger\varphi+\varphi^\dagger\psi\big)+\hbar g|\varphi|^4\,.
%sd
%\int d^2x~\hat{\Psi}^\dagger(x,y)\big[ -\frac{\hbar^2\nabla^2}{2m_{ph}}+E_{ph}\big]\hat{\Psi}(x,y)\nonumber\\&+\int d^2x~\hat{\Phi}^\dagger(x,y)\big[ -\frac{\hbar^2\nabla^2}{2m_{ex}}+E_{ex}\big]\hat{\Phi}(x,y)\nonumber\\&+\hbar g\int d^2x~\big[\hat{\Psi}^\dagger(x,y)\hat{\Phi}(x,y)+\hat{\Phi}^\dagger (x,y)\hat{\Psi}(x,y)\big]\,.
\end{align}
Here, the two first terms describe the free evolution of the photon
($\psi$) and exciton ($\varphi$) fields, respectively; the first term
in the second line describes the \emph{Rabi coupling} between the
photon and the exciton fields, which transfers excitations at the Rabi
frequency $\Omega$; the last term accounts for self--interaction in the
exciton field, with $\hbar g$ as the exciton--exciton interaction
strength.

%  Field operators $\hat{\Psi}^\dagger(\hat{\Phi}^\dagger)$ and  $\hat{\Psi}(\hat{\Phi})$ create and annihilate photons (excitons) at $(x,y)$. We can write them as:
%  \begin{subequations}
%  	\label{eq:ysgd4fy763723sdgwd}
%  	\begin{align}
%  	\hat{\Psi}&=\hat{a}\psi(x,y)\,,\\
%  	\hat{\Phi}&=\hat{b}\varphi(x,y)\,,
%  	\end{align}
%  \end{subequations}
%where annihilation and creation operators $\hat{a}$, $\hat{a}^\dagger$, $\hat{b}$ and $\hat{b}^\dagger$ obey standard bosonic commutation relations $[\hat{a},\hat{a}^\dagger]=1$ and $[\hat{b},\hat{b}^\dagger]=1$. Adding Eqs.~(\ref{eq:ysgd4fy763723sdgwd}) in Eq.~(\ref{eq:jhyh877hehyi8o3}) we have

\subsection{Main Equations}

Based on the Hamiltonian in Eq.~(\ref{eq:jhyh877hehyi8o3}), the
dynamics of 2D interacting polaritons is described by a set of coupled
Schr\"odinger equation and Gross Pitaevskii equation for the
photon~$\psi$ and exciton~$\varphi$ fields, respectively:
\begin{equation}
  \label{eq:Mon22May104821BST2017}
  i\hbar\partial_t
  \begin{pmatrix}
    \psi(x,y,t)\\
    \varphi(x,y,t)
  \end{pmatrix}
  =
  \mathcal{L}
  \begin{pmatrix}
    \psi(x,y,t)\\
    \varphi(x,y,t)
  \end{pmatrix}\,,
\end{equation}
where 
\begin{align}\label{eq:Mon22May104821BST2018}
\mathcal{L}=\begin{pmatrix}
-\frac{\hbar^2\nabla^2}{2m_{ph}}+E_{ph} & \hbar \Omega \\
\hbar \Omega & -\frac{\hbar^2\nabla^2}{2m_{ex}}+E_{ex}+\hbar g|\varphi|^2
\end{pmatrix}\,.
\end{align}
Here, $m_{ex}=(m_e+m_h)m_0$ stands for the exciton mass and
$m_{ph}\approx n\pi \hbar/(c l_c)=(n^2/c^2)(E_{x}+\delta)$ is the
effective photon mass, with $E_{ph}\approx \hbar c \pi/(n l_c)$ where
$l_c$ is the microcavity length. $E_{ex}$ is the exciton energy that
is given by $E_{ex}=\SI{13.6}{\electronvolt}\frac{\mu}{m_0 n}$, where
$\mu$ is the reduced effective mass of the exciton and $m_0$ is the
bare electron mass. For later calculations, we may need equations in a
rescaled form, then one can rescale the units of time and length as
$t\rightarrow t\Omega$, $x\rightarrow x/\xi$, $y \rightarrow y/\xi$,
$k_{x}\rightarrow k_x\xi$, and $k_{y}\rightarrow k_y\xi$ where
$\xi\equiv\sqrt{\hbar/\Omega m_{ph}}$.

Theoretical works \cite{Tassone99} have estimated the self-interaction
strength as $\hbar g\approxeq 6E_x a_B^2$, where $a_B$ is the
exciton Bohr radius. Calculating $\hbar g/(\hbar^2/2m_p)$, with
$m_p\approx 10^{-4}m_0$ as the polariton mass, one finds that
polaritons are in the regime of weak interactions. However,
experimental evidence \cite{Sun17a,Rosenberg18} has recently suggested
that the interaction constant $\hbar g$ could be of the order of
several $\SI{}{\milli\electronvolt\micro\meter}$, which puts
polaritons in the regime of strong self-interaction. In this work, we
aim at considering the effect of energy detuning on the dynamics of
coupled fields with phase defects, with an emphasis on the role of the
Rabi coupling as opposed to polariton-polariton interactions. Such a
regime of vortex dynamics has been reported recently in
Ref.~\cite{domicini18a}, which seems to justify the realm of a linear
dynamics of coupled fields; therefore, hereafter, we ignore the effect
of self-interaction energy in our calculations.

\subsection{Initial states}\label{subsec:hrtuer8347209332iunjwe}

Our objective here is to study the dynamics of vortices in the regime
of strong coupling. To this aim we introduce vortices in the initial
condition as follows:
%%
%\begin{enumerate}
%\item In the photon field: a superposition of a topological charge
%  $\mathrm{TC}=1$ Gaussian with a $\mathrm{TC}=0$ Gaussian, the latter
%  centered at~$(x_c(0),y_c(0))$.
%\item In the exciton field: a $\mathrm{TC}=1$ Gaussian, with its core
%  centered at $(x_x(0),y_x(0))$.
%\end{enumerate}
%%
%This reads
%%
\begin{subequations}
	\label{eq:Mon22May120803BST2017}
	\begin{align}
	\psi_0\equiv\psi(x,y,t=0)=&\frac{e^{-(x^2+y^2)/2w^2}}{w\sqrt{\pi(w^2+|z_c(0,0,0)|^2)}}z_c(x,y,0)\,,\\
	\label{eq:Mon23May1200803BST2017}
	\varphi_0\equiv\varphi(x,y,t=0)=&\frac{e^{-(x^2+y^2)/2w^2}}{w\sqrt{\pi(w^2+|z_x(0,0,0)|^2)}}z_x(x,y,0)\,,
	\end{align}
\end{subequations}
where~$z_j(x,y,t)=x-x_j(t)+i(y-y_j(t))$ with
$j=\left\lbrace c,x\right\rbrace $. Here, each field is
normalized. The parameter $w$ control the Gaussian spot size. It is
worth mentioning that one can consider coupled fields with different
topological charge in each, however, we found out that the initial
states (\ref{eq:Mon22May120803BST2017}) are the simplest choice to
illustrate the physics we want to discuss. Also, note that an initial
condition with a vortex in the photon field only does not develop any
dynamics for the vortex core.

\section{Oscillation of the core}\label{se:jhgyhsyt84347udgi}

In this section we discuss the solutions of equation
(\ref{eq:Mon22May104821BST2017}) when $\Omega=0$. Despite the fact
that the equations are linear, we will see that the solutions are not
trivial.

 \begin{figure*}[t]
 	\begin{center}
 		\includegraphics[width=\linewidth]{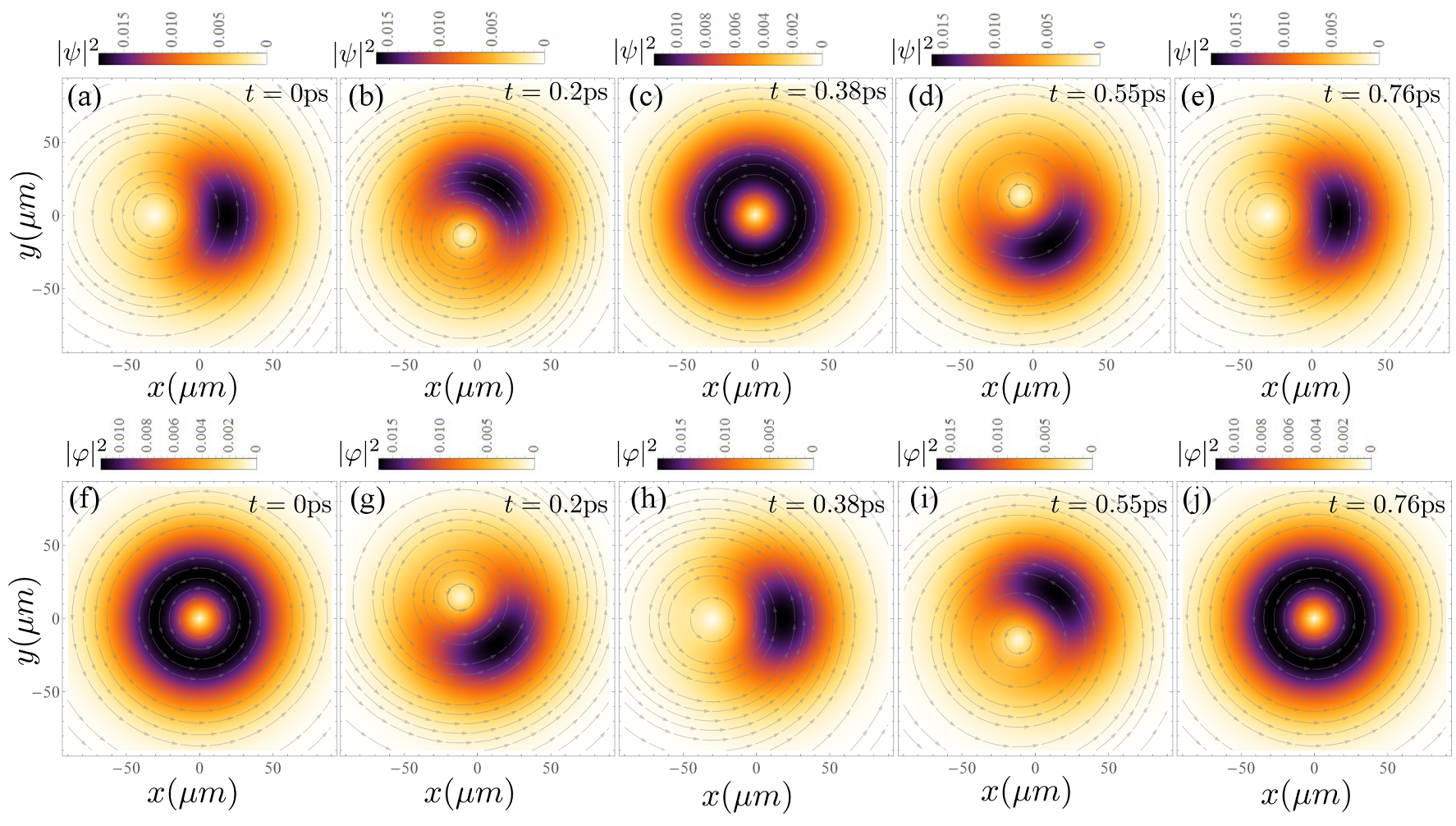}
 		\caption{Oscillation of the core in the photon (upper
                  panels) and in the exciton (lower panels) fields,
                  shown for one period of motion. The wavepacket is
                  initially normalized to one and the energy detuning
                  is zero, which implies no particle exchange between
                  the two fields, which remain normalized in
                  time. Each fields carries a winding number of
                  $l=1$. The photon field is initially deformed by
                  displacing the core to a position far from the
                  origin, while the core in the exciton field is
                  located at the origin. For the dynamics, the local
                  density oscillates similarly. The core does not move
                  with a constant speed. We also show the stream lines
                  of superfluid velocity for each field, as circles
                  centered on the vortex core position. Here, we used:
                  $\hbar \Omega=2.75\mathrm{meV}$, $a=1.2$,
                  $w=25\mathrm{\mu m}$.  }
 		\label{fig:rytuwteitrgywey89382hj}
 	\end{center}
 \end{figure*}  

\subsection{Approximate Solution} 

There is no closed-form solution for
Eq.~(\ref{eq:Mon22May104821BST2017}) even in the absence of a
nonlinear term. However, one can find approximate solutions through a
series expansion, for example, by employing spectral methods
\cite{Num} or homotopy analysis methods \cite{Shi12}. In this text, we
use a spectral method, with the Hermite $H_n(x)$ functions as basis
functions; thus, we expand the wavefunctions $\psi(x,y)$ and
$\varphi(x,y)$ as:
\begin{subequations}
\begin{align}
\psi(x,y)&=\sum_{n,m=0}u_{n,m}(x,y)a_{n,m}(t)\,,\\
\varphi(x,y)&=\sum_{n,m=0}u_{n,m}(x,y)b_{n,m}(t)\,,
\end{align}
\end{subequations}
where we introduce the basis functions $u_{n,m}(x,y)$ as:
\begin{align}
u_{n,m}(x,y)&=c_{n,m}H_n(x/w)H_m(y/w)e^{-(x^2+y^2)/(2w^2)}\,,
\end{align}
with
$c_{n,m}=\frac{1}{w\sqrt{\pi2^{n+m}\Gamma(n+1)\Gamma(m+1)}}$. Here,
$\Gamma(n)$ stands for the gamma function. The coefficients $a_{n,m}$ and
$b_{n,m}$ satisfy the following initial-value differential
equations:
\begin{widetext}
  \begin{subequations}
    \begin{align}
      i\hbar\frac{da_{n,m}}{dt}&=-\frac{\hbar^2}{2m_{c}w^2}\big(-(1+n+m)a_{n,m}+\frac{\sqrt{(n+1)(n+2)}}{2}a_{n+2,m}+\frac{\sqrt{(m+1)(m+2)}}{2}a_{n,m+2}\nonumber\\&+\sqrt{n(n-1)}a_{n-2,m}+\sqrt{m(m-1)}a_{n,m-2}\big)+\hbar \Omega b_{n,m}+E_ca_{n,m}\,,\\
      i\hbar\frac{db_{n,m}}{dt}&=-\frac{\hbar^2}{2m_{x}w^2}\big(-(1+n+m)b_{n,m}+\frac{\sqrt{(n+1)(n+2)}}{2}b_{n+2,m}+\frac{\sqrt{(m+1)(m+2)}}{2}b_{n,m+2}\nonumber\\&+\sqrt{n(n-1)}b_{n-2,m}+\sqrt{m(m-1)}b_{n,m-2}\big)+\hbar \Omega a_{n,m}+E_xb_{n,m}\,.
    \end{align}
  \end{subequations}
\end{widetext}
These provide an infinite hierarchy of equations, coupled by the Rabi
and kinetic energy terms. One way to truncate this hierarchy is to
note that for $w^2\gg\xi^2$, only those $a_{n,m}$ and $b_{n,m}$ that
have nonzero initial values are several orders larger than the other
coefficients with zero initial values; it is the case when $w$ is of
the order of several micrometers, which is a typical estimate found in
experimental reports \cite{domicini18a}. Then, we can approximate the
wavefunctions as:
\begin{subequations}\label{eq:juhgy766489883jgf}
	\begin{align}
	\psi(x,y,t)&\approxeq a_{00}(t)u_{00}+a_{1,0}u_{1,0}+a_{0,1}u_{0,1}\,,\\
	 \varphi(x,y,t)&\approxeq b_{00}(t)u_{00}+b_{1,0}u_{1,0}+b_{0,1}u_{0,1}\,.
	\end{align}
\end{subequations}

The equations for $a_{0,0}(t)$, $a_{0,1}(t)$, $a_{1,0}$, $b_{0,0}(t)$,
$b_{0,1}(t)$ and $b_{1,0}(t)$ are given in Appendix
\ref{app:jfhruity854ierh9403}.

%It is also possible to provide another set of solutions, as far as we consider  initial states in Eqs.~(\ref{eq:Mon22May120803BST2017}) and the condition for $w^2\gg\xi^2$ is satisfied; as our system is linear one expects the solutions that are combinations of $u_0$ and $u_1$ with  
%%
%\begin{align}
%u_l(r,\phi)=\frac{1}{\sqrt{\pi \Gamma(|l|+1)}}\frac{1}{w^{|l|+1}}r^{|l|}e^{-r^2/{2w^2}}e^{i\phi l}\,,
%\end{align}
%%
%which is given in polar coordinate $r=\sqrt{x^2+y^2}$ and $\phi=\arg(x+iy)$; then our solutions are:
%%
%\begin{subequations}
%	\label{eq:gduytgeuye639823}
%	\begin{align}
%	\psi\approxeq&a_0(t)u_0(x,y)+a_1(t)u_1(x,y)\,,\\
%	\varphi\approxeq&b_0(t)u_0(x,y)+b_1(t)u_1(x,y)\,.
%	\end{align}
%\end{subequations}
%% 
%we also provides solutions for $a_0(t)$, $a_1(t)$, $b_{0}(t)$, and $b_1(t)$ in appendix \ref{app:jfhruity854ierh9403}. Here then after, we use approximate solutions in Eq.~(\ref{eq:juhgy766489883jgf}).
%
\begin{figure}[t]
	\begin{center}
		\includegraphics[width=\linewidth]{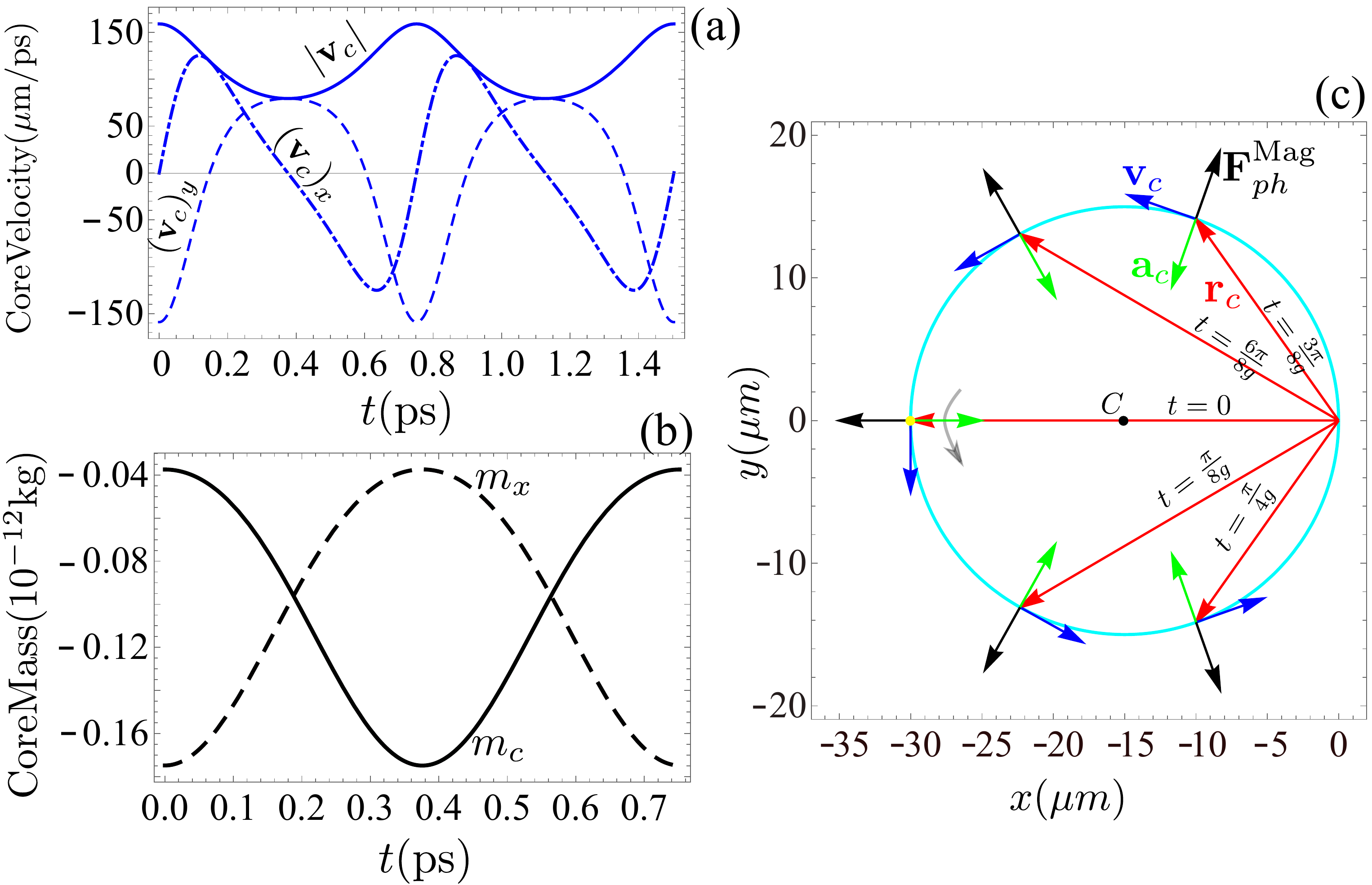}
		\caption{(a): absolute value of core velocity
                  $|\mathbf{v}_c|$ (solid line in blue) in the photon
                  field at zero detuning. We also show its components
                  $(\mathbf{v}_c)_x$ (dashed line) and
                  $(\mathbf{v}_c)_y$ (dot--dashed line). The core is
                  faster when it is the farthest from the origin. (b):
                  trajectory (in cyan) of the core (yellow point) in
                  the photon field. Position vector (in red), velocity
                  vector (in blue), acceleration vector (in green),
                  and Magnus force vector (in black) are shown for
                  different times. The acceleration vector points
                  toward the center $C$ (black point). The Magnus
                  force points in the opposite direction. (c) shows
                  the core mass in the photon (solid line) and exciton
                  (dashed line) fields. The core is heavier at the
                  origin. For numerical calculation, we used the same
                  parameter as
                  Fig.~(\ref{fig:rytuwteitrgywey89382hj}).}
		\label{fig:rytuwasateitrgywey89382hj}
	\end{center}
\end{figure}  
\subsection{Dynamics of the core}
Examples of solutions for zero detuning are shown in
Fig.~\ref{fig:rytuwteitrgywey89382hj}. The cores are initially located
at $(x_c(0)=-a w,y_c(0)=0)$ in the photon field and at
$x_x(0)=0,y_x(0)=0$ in the exciton field. While they move periodically
on their orbits, the background densities oscillate similarly. Here,
the fields are initially normalized and remain so in time. This
implies that there is no particle exchange between the coupled fields,
however, background densities are oscillating while the core
moves. The motion of the core is induced by the presence at a distance
of the other core, by the local density of the other field, and by
the Rabi coupling. The combined effect is that the vortex core moves
with a non-constant speed.

We now explain the dynamics based on our main equations. One notes
that the core in each field never disappears completely, which implies
that the solutions can be written as the product of a background
density and a complex time-dependent function
$z_j(t)=x-x_j(t)+i(y-y_j(t))$ for depleted points in the fields, as we
have:
\begin{subequations}\label{eq:hyugjytrhg675489746}
\begin{align}
\psi&=\rho_{ph}(x,y,t)e^{i\phi_{ph}(x,y,t)}z_c(t)\,,\\
\varphi&=\rho_{ex}(x,y,t)e^{i\phi_{ex}(x,y,t)}z_x(t)\,.
\end{align}
\end{subequations}
By adding them in Eq.~(\ref{eq:Mon22May104821BST2017}) one can find
two central equations for the core velocities:
\begin{subequations}\label{eq:juhygt67587yhftgd36}
\begin{align}
\mathbf{v}_c&=\frac{\hbar}{m_{ph}}\big( \vec\nabla\phi_{ph}-\hat{k}\times \vec \nabla\ln \rho_{ph}\big)|_{x=x_c,y=y_c}\nonumber\\&+\Omega\frac{\rho_{ex}}{\rho_{ph}}\big( \hat{j}\Delta \mathbf{r}\cdot\hat{S}_++\hat{i}~\hat{k}\times\Delta \mathbf{r}\cdot\hat{S}_+\big)\,,\\
\mathbf{v}_x&=\frac{\hbar}{m_{ex}}\big( \vec\nabla\phi_{ex}-\hat{k}\times \vec \nabla\ln \rho_{ex}\big)|_{x=x_x,y=y_x}\nonumber\\&-\Omega\frac{\rho_{ph}}{\rho_{ex}}\big( \hat{j}\Delta \mathbf{r}\cdot\hat{S}_-+\hat{i}~\hat{k}\times\Delta \mathbf{r}\cdot\hat{S}_-\big)\,,
\end{align}
\end{subequations}
where $\Delta \mathbf{r}=\mathbf{r}_c-\mathbf{r}_x$ gives the relative
vector-position of the cores, and
$\hat{S}_\pm=\cos(\phi_{ph}-\phi_{ex})\hat{i}\pm\sin(\phi_{ph}-\phi_{ex})\hat{j}$.These
equations have the same mathematical form even in the presence of
self-interactions and external potentials, although with a potential
the dynamics of the cores take a different nature. Also, we note that
there are three factors that determine the variation of velocities in
time; one factor depends on the distance between the cores in each
field which is followed by the Rabi frequency $\Omega$. The other two
factors depend on the gradient of the local phase and density of each
field; these two factors come from the kinetic energy through a
dependency on $\hbar/m_{ph,ex}$. Comparing $\hbar /m_cw$ and $\Omega w$ to
approximate the order of the factors that define the cores velocity,
we assume, based on experimental evidences, that $\Omega\gg \hbar/m_cw^2$,
which implies that the Rabi--induced velocity is the dominating factor
for the motion of the vortex core.  For $\Omega=0$, each field turns to a
free field, and there is no motion of the core. In our case, there is
no external potential, and the motion of the core is mediated only by
the interplay between the Rabi coupling and the topology of the
deformed field when a core is displaced to a point far from the center of
the field. As the core is moving on a curved path, it has an angular
velocity $\vec\omega$ which is given by:
\begin{align}
\vec\omega_{j}=\frac{\mathbf{r}_j\times\mathbf{v}_{j}}{x_j^2+y_j^2}\,,
\end{align}
with $j=\left\lbrace c,x\right\rbrace $ and $\mathbf{r}_j$ denoting
the position vector of the core in the photon ($j=c$) and in the
exciton ($j=x$) fields.
 
At this stage, we have shown that a core has both a linear and angular
velocity, and its kinematic (description of the motion itself with no
reference to the underlying forces) is thus transparent. To address
the kinetics (descriptions of the causes of motions) of the core,
however, we also need to refer to the force that acts on it. In a
classical fluid, the circulation of a vortex can influence any other
vortex at a distance, which results in an effective force
perpendicular to the velocity of the vortex
core\cite{Calderaro17}. Bringing this picture to a quantum fluid, it
was shown that such a force is connected to the gradient of the total
energy of the system for a BEC in an harmonic
trap\cite{Jackson99,Donadello14}, and ambiguously for two unbounded
coupled BECs\cite{Calderaro17} in the nonlinear regime. For polaritons
in the linear regime, one can show that the predominant energy-scale
in the system is the Rabi energy, defined as:
\begin{align}\label{eq:kjuhy6554tgftrsyd7e6}
E_R(x_c,y_c,x_x,y_x)\equiv 2\hbar \Omega\mathrm{Re}\langle\psi|\varphi \rangle\,,
\end{align}
and this depends on the positions of the vortex cores. Taking the
gradient with respect to the core positions, one can introduce a force
through $\mathbf{F}^{\mathrm{R}}\equiv-\vec\nabla E_R$. For a core in
the photon field, the $x$ and $y$ components of this force have a
linear dependency on $x_x,y_x$ through an equation like
$\Omega(\alpha_1(t)x_x+\alpha_2(t)y_x)$, while for a core in the exciton
field, it behaves like $\Omega(\beta_1(t)x_c+\beta_2(t)y_c)$, where
$\alpha_i$ and $\beta_i$ are some real coefficients depending on
the parameters of the system. On the other hand, there would be a Magnus--like
force that depends on the velocity of the vortex core. To find this,
we multiply velocities in Eqs.~(\ref{eq:juhygt67587yhftgd36}) from the
left respectively by $-2\pi \hbar\rho_{ph}\hat{k}$ and
$-2\pi\hbar\rho_{ex}\hat{k}$, and we find:
\begin{subequations}\label{eq:hre89w837uuwiey}
	\begin{align}
\mathbf{F}_{ph}^{\mathrm{Mag}}+\mathbf{F}_{ph}^{\mathrm{Rabi}}&=\mathbf{F}_{ph}^{\mathrm{Kin}}\,,\\
\mathbf{F}_{ex}^{\mathrm{Mag}}+\mathbf{F}_{ex}^{\mathrm{Rabi}}&=\mathbf{F}_{ex}^{\mathrm{Kin}}\,,
	\end{align}
\end{subequations}
where we introduce the following expressions; for the core in the photon field:
\begin{subequations}
	\begin{align}
	\mathbf{F}_{ph}^{\mathrm{Mag}}&=-\rho_{ph}\mathbf{K}_{ph}\times (\mathbf{v}_c-\frac{\hbar}{m_{ph}}\vec\nabla\phi_{ph})\,,\\
		\mathbf{F}_{ph}^{\mathrm{Kin}}&=-\frac{\hbar\rho_{ph}}{m_{ph}}\mathbf{K}_{ph}\times (\hat{k}\times \vec \nabla\ln \rho_{ph})\,,\\
		\mathbf{F}_{ph}^{\mathrm{Rabi}}&=\Omega\rho_{ex}\mathbf{K}_{ph}\times\big( \hat{j}\Delta \mathbf{r}\cdot\hat{S}_-+\hat{i}~\hat{k}\times\Delta \mathbf{r}\cdot\hat{S}_-\big)\,,
	\end{align}
\end{subequations}
and for the core in the exciton field:
\begin{subequations}
	\begin{align}
	\mathbf{F}_{ex}^{\mathrm{Mag}}&=-\rho_{ex}\mathbf{K}_{ex}\times (\mathbf{v}_x-\frac{\hbar}{m_{ex}}\vec\nabla\phi_{ex})\,,\\
	\mathbf{F}_{ex}^{\mathrm{Kin}}&=-\frac{\hbar\rho_{ex}}{m_{ex}}\mathbf{K}_{ex}\times (-\hat{k}\times \vec \nabla\ln \rho_{ex})\,,\\
	\mathbf{F}_{ex}^{\mathrm{Rabi}}&=-\Omega\rho_{ph}\mathbf{K}_{ex}\times\big( \hat{j}\Delta \mathbf{r}\cdot\hat{S}_++\hat{i}~\hat{k}\times\Delta \mathbf{r}\cdot\hat{S}_+\big)\,,
	\end{align}
\end{subequations}
where $\mathbf{K}_{j}=2\pi \hbar \hat{k}$. Here, we have a balance
between the Magnus force $\mathbf{F}^{\mathrm{Mag}}_j$ added to a Rabi
mediated force $\mathbf{F}^{\mathrm{Rabi}}_j$ and a gradient force
$\mathbf{F}^{\mathrm{Kin}}_j$. The latter force has an origin in
quantum kinetic energy or zero point energy, and is related to the
quantum pressure. Examining the orbital motion of the core, as it is
shown in Fig.~(\ref{fig:rytuwteitrgywey89382hj}), one can conclude
that a possible force that makes the core drift should depend on the
relative distance between the cores, as when $x_c(0)=x_x(0)$ and
$y_c(0)=x_c(0)$ there is no motion for the core. This implies that any
effective force $\mathbf{F}^\mathrm{R}$ related to the gradient of the
Rabi energy does not reproduce the expected dynamics of the vortex
core in our binary system.
\begin{figure}[t]
	\begin{center}
		\includegraphics[width=\linewidth]{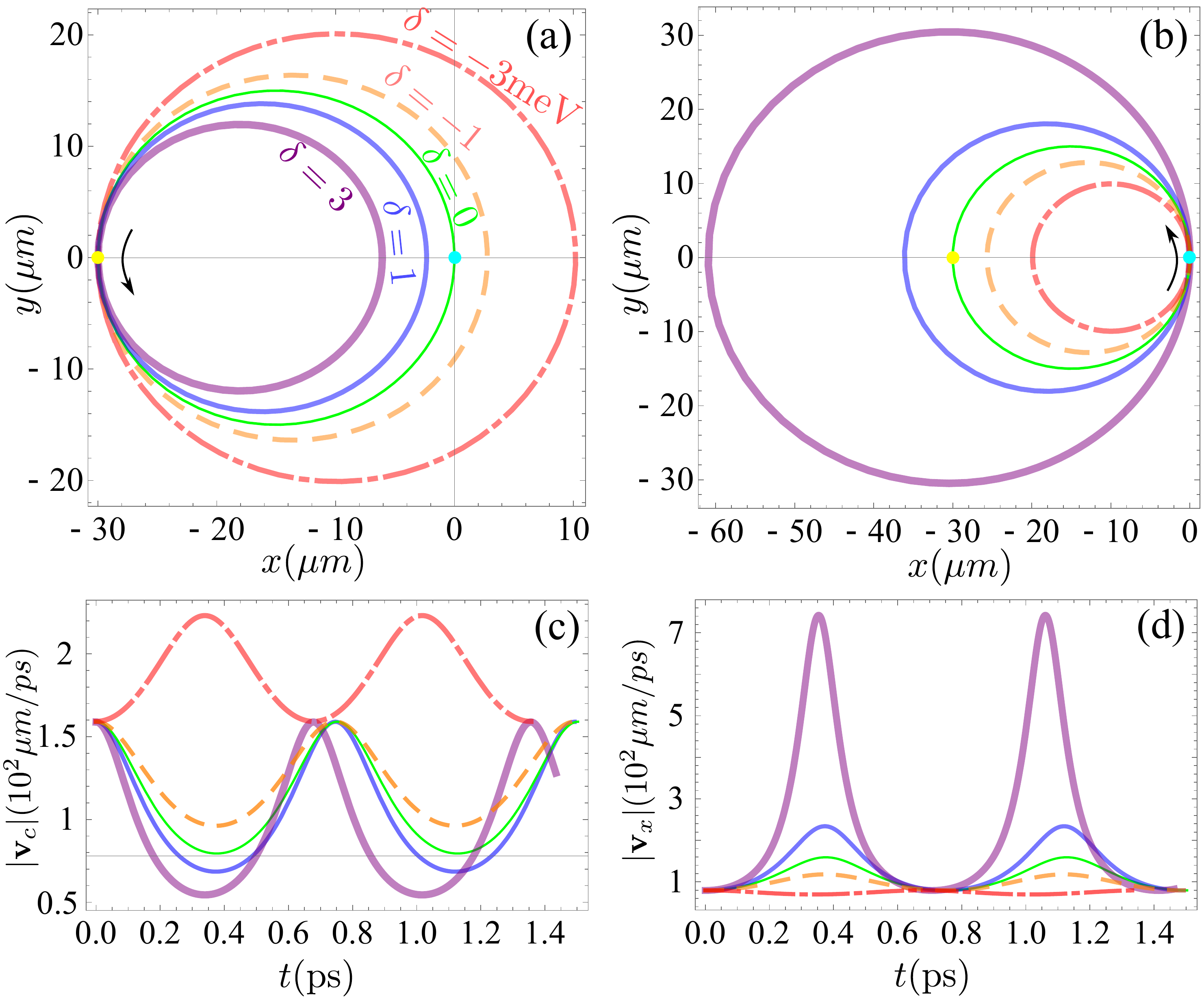}
		\caption{(a) and (b) show orbits for the core
                  oscillations for different energy detuning between
                  the photon and exciton fields. The core in the
                  photon field is shown as a yellow point, and in the
                  exciton field as a green point. In (a) and for
                  positive (negative) detuning, the core moves on
                  smaller (larger) orbits, and its dynamical
                  properties are different. It opposite happens for
                  the core in the exciton field. The amplitude of
                  velocity ($|\mathbf{v}_c|$ and $|\mathbf{v}_x|$) are
                  shown in (c) and (d). As the core takes larger
                  orbits, it speeds up even up to speeds larger than
                  the speed of light.}
		\label{fig:rytuwastrgywey89382hjikj}
	\end{center}
\end{figure} 

Based on the above equations for the forces, we now have a better
understanding for an effective mass associated to the core. To this
end, we note that the application of a force normal to the velocity
direction does change the linear momentum of an object, and the rate
of the change in momentum is equal to the applied force. This implies
that there would be a rotational acceleration related to the cross
product of the linear and angular velocities, which in the case of the
core motion results in
\begin{subequations}\label{eq:gftry86897209iwuyue}
	\begin{align}
	\mathbf{F}_c&\equiv m_c\vec\omega_c\times\mathbf{v}_c=-\rho_{ph}\mathbf{K}_{ph}\times\mathbf{v}_c\,,\\
	\mathbf{F}_x&\equiv m_x\vec\omega_x\times\mathbf{v}_x=-\rho_{ex}\mathbf{K}_{ex}\times\mathbf{v}_x\,,
	\end{align}
\end{subequations}
where $m_c$ and $m_x$ are the core masses in the photon and exciton
fields, respectively, and we have:
\begin{subequations}
	\begin{align}
	m_c&=-\frac{2\pi \hbar\rho_{ph}(x_c,y_c)}{\omega_c}\,,\\
	m_x&=-\frac{2\pi \hbar\rho_{ex}(x_x,y_x)}{\omega_x}\,,
	\end{align}
\end{subequations}
which depend explicitly on the angular velocity of the cores. As there
is no dependency on the Rabi frequency, one can be hopeful that even
in the presence of self-interaction or external potentials, the
equations for the masses $m_c$ and $m_x$ will have the same
mathematical form, and that any effects associated with interactions
and potential will be collected in angular frequencies of the cores
and ambient densities.
    
Now we have a complete description for the core dynamics with its
position, velocity, mass and force associated to the core. In
Fig.~\ref{fig:rytuwasateitrgywey89382hj}, we show examples of the
dynamical variables for zero detuning and for the photon field. The
core is initially located at $(-aw,0)$, as is indicated by the yellow
point in Fig.~\ref{fig:rytuwasateitrgywey89382hj}(a). While the core
moves counter-clockwise, its speed $|\mathbf{v}_c|$ decreases to a
minimum, when the core finds itself at the origin, and then the core
speeds up as it approaches to its initial position. With such a
variation in its speed, the core decelerates in the first half of its
motion and then accelerates in the rest of its motion. The
acceleration vector is pointed toward a point $C$ inside the orbit,
while the Magnus force is pointed in the opposite direction. This
implies a negative mass for the core, which is shown in
Fig.~\ref{fig:rytuwasateitrgywey89382hj}(b) for one period of
motion. The core is lighter when it take its position at the periphery
of the local density. Since in our system, there is no external
potential, such a variation in the inertial mass of the core
originates exclusively from the binary nature of the
fields. Certainly, with the presence of an external potential and/or
self-interactions, the core mass would behave differently, depending
on the interplay between the Rabi energy and the other energy scales
of the system.

Now we turn to consider the effect of the energy detuning. We first
study the orbital motions of the cores for different detunings, which
are shown in Fig.~\ref{fig:rytuwastrgywey89382hjikj}(a) and (b) for
the photon and exciton fields, respectively. With a nonzero detuning,
the cores move on different orbits, while for a zero detuning, they
orbit on the same path. Negative and positive detunings have different
effects, as the cores can move on smaller or larger orbits.  Figures
\ref{fig:rytuwastrgywey89382hjikj}(c) and (d) show the corresponding
amplitudes of velocities: $|\mathbf{v}_c|$ and $|\mathbf{v}_x|$. As
the detuning has opposite effects in the photon and exciton fields, it
decreases or increases the core speeds. It seems, as predicted in \cite{domicini18a}, that the core speed
could increase even to value larger than the speed of light
$c=\SI{3e2}{\micro\meter\per\pico\second}$, and at the same time,
there would be a decrease of its mass. An angular momentum (AM)
polariton state
$|\Psi\rangle=\alpha|\psi\rangle_{l_c}+\beta|\varphi\rangle_{l_x}$,
where $|\psi\rangle_{l_c}$ and $|\varphi\rangle_{l_x}$, respectively,
indicate photon and exciton states with different topological charges
$l_c$ and $l_x$, will propagate with a group velocity that depends on
the topology of the fields. One could expect that by increasing and/or
decreasing the detuning and equivalently manipulating the core speed,
the group velocity will increase or decrease correspondingly. This
controlling over group velocity has practical importance in quantum
memory\cite{Zhong17}.

\section{Angular momentum and Energy Contents}\label{se:juyftg73493jew92302}
A wavepacket with a phase defect that results in a rotation of the
fluid around a vortex core, has a nonzero value of angular
momentum. Namely, the average angular momentum of the wavepacket
$\Psi$ is given by
\begin{align}
\langle \hat{L}_z\rangle&=-i\hbar \int r dr d\phi \langle \Psi|\frac{\partial}{\partial\phi}\Psi\rangle\,,
\end{align}
with $r=\sqrt{x^2+y^2}$ and $\phi=\arg(x+iy)$ in polar
coordinates. For the initial wavepackets in
Eqs.~(\ref{eq:Mon22May120803BST2017}), one can find
$\langle \hat{L}_z\rangle_{ph}(t=0)=\hbar w^2/(w^2+|z_c|^2)$, and
$\langle \hat{L}_z\rangle_{ex}(t=0)=\hbar w^2/(w^2+|z_x|^2)$,
respectively, for the photon and exciton fields; then, with
$z_c\neq z_x$, corresponding to placing the cores in different points
in space, there is a nonzero initial imbalance for the angular
momentum
$\Delta l\equiv \langle \hat{L}_z\rangle_{ph}-\langle
\hat{L}_z\rangle_{ex}\neq0$. This also changes the energy content of
the coupled fields. To study this, we first note that the kinetic
energy is of the order of $\hbar \Omega/2w^2$ for the photon field and of
$(m_{ph}/m_{ex})\hbar \Omega/2w^2$ for the exciton field; and that for $w$
equal to several $\xi$, these are at least two orders less than the
Rabi energy which is proportional to $\hbar \Omega$. This means that one
can neglect the kinetic energy for $w$ in the range of several
$\xi$. For zero detuning, the Rabi energy is a constant of motion and
is given at $t=0$ by
$E_R(t=0)=2\hbar
\Omega(w^2+\mathbf{r}_c\cdot\mathbf{r}_x)/\sqrt{(w^2+|z_c|^2)(w^2+|z_x|^2)}$. The
configuration with $\mathbf{r}_c=\mathbf{r}_x$ is unstable to a
displacement of the cores, since by increasing the distance between the
two cores, the energy of the system decreases correspondingly. One can
deduce that states with cores at a distance are more favorable than
$\mathbf{r}_c=\mathbf{r}_x$, corresponding to fields being at rest,
and this decreasing in energy as well as inducing the nonzero
$\Delta l$, will be realized through the rotation of the vortex cores.
\begin{figure}[t]
	\begin{center}
		\includegraphics[width=\linewidth]{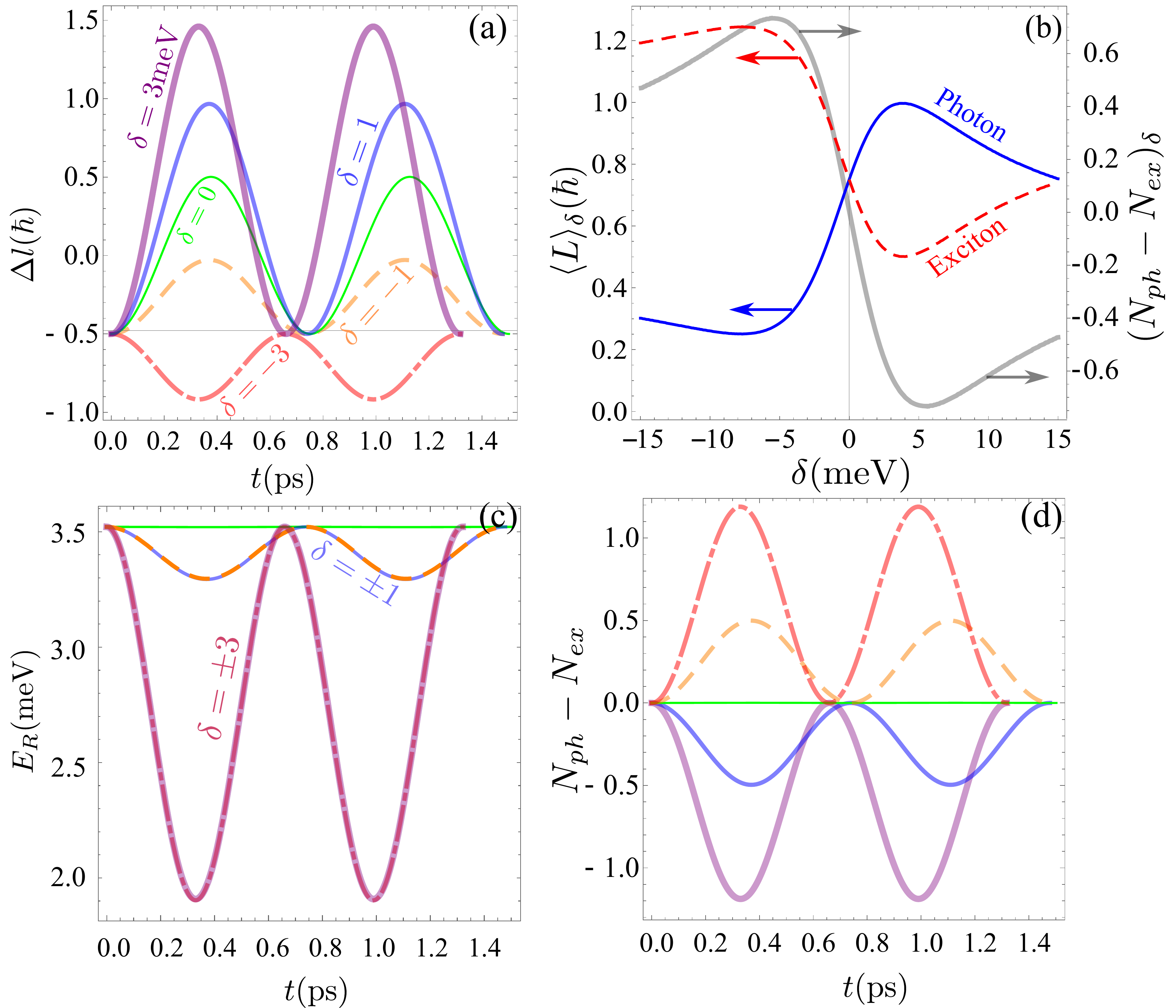}
		\caption{(a) shows the imbalance in angular momentum:
                  $\Delta l=\langle L_z\rangle_{ph}-\langle
                  L_z\rangle_{ex}$. For positive detuning, the angular
                  momentum of the photon field is larger. We see
                  variations in the period of oscillations with
                  detuning. It is shown for $\delta=3\mathrm{meV}$ in
                  purple, $\delta=1\mathrm{meV}$ in blue,
                  $\delta=0\mathrm{meV}$ in green,
                  $\delta=-1\mathrm{meV}$ in orange, and
                  $\delta=-3\mathrm{meV}$ in red. (b) The mean value
                  of the angular momentum for the photon (blue--solid
                  line) and exciton (red--dashed line) fields. For
                  negative detunings, the angular momentum is carried
                  mostly by the photon field. The mean value of the
                  population imbalance is also shown (in dark--thick
                  solid line). In (c) we present the variations of the
                  Rabi energy in time for different detuning. In (d)
                  we show population imbalance.}
		\label{fig:ry87wastrgywey89382hjikj}
	\end{center}
\end{figure} 

We now describe the time evolution of the energy and angular momentum
in Fig.~\ref{fig:ry87wastrgywey89382hjikj} for some negative and
positive detuning. One can see in
Fig.~\ref{fig:ry87wastrgywey89382hjikj}(a) the time evolution of
$\Delta l$. There will be an exchange of the angular momentum between
the components of the coupled fields, which is periodic in time, as we
expect from the Rabi nature of the coupling. By placing the photon
vortex core at a distance to the exciton vortex core, there exists a
nonzero (here, negative) initial imbalance in angular momentum, as the
photon part has initially less rotational content than the exciton
part, and since the total angular momentum should be constant, this
imbalance triggers oscillations in the angular momentum. Depending on
the energy detuning, there will be oscillations with different periods
and behaviors. First we see that by increasing the detuning, there is
a decrease in the period of oscillations; then, we see that for
positive detuning, there will be longer time intervals with positive
imbalance in angular momentum, i.e., the photon field has more angular
content in average. On the other hand, with a negative detuning, there
would be a state with always $\Delta l<0$ in time, which implies that
the rotational content of the photon field is always less than the
exciton field. In this case, the core will be faster in the photon
field, as it is being pushed to the periphery of the beam, while the
core in the exciton field is slower (see
Fig.~\ref{fig:rytuwastrgywey89382hjikj}). For an object with a
positive mass, one expects that increasing the angular momentum will
make it go faster, however, in the case of a vortex core, the opposite
happens. Actually, a core is an absence of matter, and it has a
negative mass, which results in it being pushed toward regions of low
density. It is shown in part (d) in
Fig~\ref{fig:ry87wastrgywey89382hjikj}, that for a negative detuning,
the photon field has more population, and indeed the field has a
higher density in the central region, which finally pushes the core,
with its negative mass, to the lower density region, i.e., at the
periphery of the wavepacket. In part (b), we show the mean value of
the angular momentum (right $y$ axis) and of the population imbalance
(left $y$ axis). For positive detuning, the population is concentrated
in the exciton side and at the same time the photon field has more
angular momentum. The reverse happens for negative detuning. One notes
that, while the mean value of the population imbalance is symmetric
with respect to detuning variations, it is not the case for the
angular momentum. It makes the negative detuning a special case for
the dynamics of the vortex core. Indeed, if we imagine
$\mathbf{r}_c=(-aw,0)$ and $\mathbf{r}_x=(0,0)$, one can show that for
$\delta\approx-\hbar \Omega(a^2 w^2+\sqrt{1+a^2})/w^2\sqrt{1+a^2}$, there
is no evolution of the angular momentum in time, i.e.,
$\Delta l=\mathrm{cons.}$. The total energy of the system
$E_T\approx E_R(t)+\delta N_{ph}(t)$ also changes with detuning,
although it is a constant of motion, i.e., it does not change with
time. Increasing $|\Delta \mathbf{r}|$ and applying detuning, one can
set the energy of the system to a desired value. We present the time
evolution of the Rabi energy in part (c). Here, as we mentioned
before, there is no oscillation in time for zero detuning, due to the
zero imbalance in population and the normalization of the
wavepackets. Nonzero detuning will however induce oscillations of the
Rabi energy. To understand this, we first note that
$\partial_tE_R\propto-\delta \mathrm{Im}\langle \psi|\varphi\rangle$
and
$\partial_t\mathrm{Im}\langle \psi|\varphi\rangle\propto\delta E_R$;
then as $E_R(t=0)$ is nonzero, and any nonzero detuning will affect
$\langle\psi|\varphi\rangle$, and therefore there is an oscillation in
the relative phase $\arg[\langle\psi|\varphi\rangle]$, which drives
the population imbalance; hence, a nonzero detuning induces
self-trapping for both the population and the angular content of the
coupled fields.

%$\langle \hat{L}_z\rangle_{ph}(t=0)=\hbar w^2/(w^2+|z_c|^2)$, and $\langle \hat{L}_z\rangle_{ex}(t=0)=\hbar w^2/(w^2+|z_x|^2)$; 

\section{conclusion}\label{sec:juhy76tdfdr43gshwqw00}

In conclusion, we have studied the dynamics of vortices in coupled
exciton--photon fields, where the binary nature of the coupling fields
sets the vortex cores in motion. We found out that the vortex core
behaves like particles that can be described by their position,
velocity and mass. We identified the later based on the Magnus
force. Then, we showed the effect of energy detuning to control the
dynamics of the vortex core. This provides a control over both
the kinematic (positions \& velocities) and kinetics (energy \& angular
momentum) of the field plus the core positioned in the field. This
control over the vortex dynamics, that is special for polaritonic
system, can have applications in many fields related to angular
momentum in matter and light, in particular, for memory reading and writing
\cite{Ma17,Zhong17,Sigurdsson14} and information processing with
angular momentum\cite{Wang2012}. It also could be used for
manipulating and maintaining vortices in superconductor
phases\cite{Veshchunov16,blatter_vortices_1994}.

\appendix
\section{Details for Equations \ref{eq:juhgy766489883jgf}}\label{app:jfhruity854ierh9403}
We mentioned in the main text that the equations of motions for
$a_{n,m}$ and $b_{n,m}$ provide a general hierarchy. To solve such
systems of ordinary equations, one needs to truncate them based on
some criterion. We found that when $w$ is several orders larger than
$\xi$, we can approximate the solutions by keeping those terms that
are initially nonzero, while removing other terms due to their small
values. Our initial conditions in
subsection~\ref{subsec:hrtuer8347209332iunjwe} give the following
values for nonzero coefficients in $t=0$:
\begin{figure}[b]
	\begin{center}
		\includegraphics[width=\linewidth]{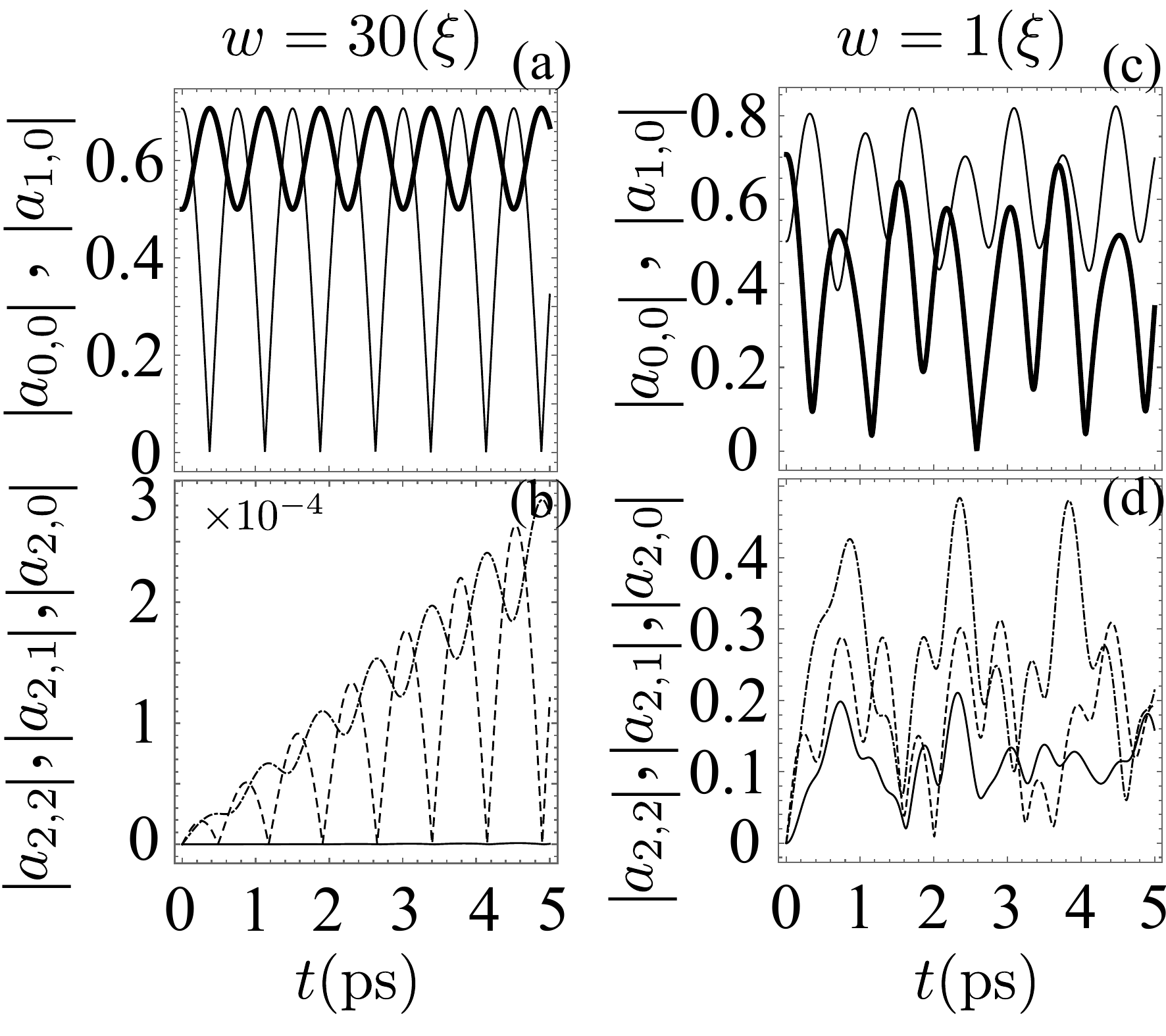}
		\caption{Variations of $|a_{n,m}|$ with $0\leq
                  m,n\leq2$ with respect to time, for two spot sizes
                  of $w$. For $w=30\xi$ in the left panels, we can keep
                  $a_{0,0}$ (solid line) and $|a_{1,0}|=|a_{0,1}|$
                  (thick solid line), while other $a_{n,m}$ are four
                  orders smaller that $a_{0,0}$, $a_{1,0}=a_{0,1}$. It
                  is shown in (b) for $|a_{2,2}|$ (solid line),
                  $|a_{0,2}|=|a_{2,0}|$ (dashed line), and
                  $|a_{1,2}|=|a_{2,1}|$ (dot--dashed line). Right
                  panels show $|a_{n,m}|$ for $w=\xi$, where all
                  coefficients are in the same order. }
		\label{fig:rytuwasateitrgywey89382hjikj}
	\end{center}
\end{figure}  
\begin{subequations}
	\begin{align}
	a_{0,0}(0)&=\frac{-(x_c+iy_c)}{\sqrt{w^2+|z_c(0,0,0)|^2}}\,,\\
	a_{0,1}(0)&=\frac{w\sqrt{2}}{2\sqrt{w^2+|z_c(0,0,0)|^2}}\,,\\
	a_{1,0}(0)&=\frac{iw\sqrt{2}}{2\sqrt{w^2+|z_c(0,0,0)|^2}}\,,\\
	b_{0,0}(0)&=\frac{-(x_x+iy_x)}{\sqrt{w^2+|z_x(0,0,0)|^2}}\,,\\
	b_{0,1}(0)&=\frac{w\sqrt{2}}{2\sqrt{w^2+|z_x(0,0,0)|^2}}\,,\\
	b_{1,0}(0)&=\frac{iw\sqrt{2}}{2\sqrt{w^2+|z_x(0,0,0)|^2}}\,,
	\end{align}
\end{subequations}
which yield the following relations:
\begin{widetext}
\begin{subequations}
	\begin{align}
	a_{0,0}(t)&=\frac{1}{g_c}e^{-\frac{it}{4w^2}(\Omega(1+m_r)+2(E_{ph}+E_{ex}) w^2)}\big( -i(a_{0,0}(0)\Omega+2w^2(a_{0,0}(0)\frac{\delta}{\hbar}+2b_{0,0}(0)\Omega))\sin(\frac{\Omega c t}{4w^2})+a_{0,0}(0)g_c\cos(\frac{\Omega c t}{4w^2})\big)\,,\\
	a_{1,0}(t)&=\frac{1}{g_d}e^{-\frac{it}{2w^2}(\Omega(1+m_r)+2(E_{ph}+E_{ex}) w^2)}\big(-i\sin(\frac{\Omega d t}{2w^2})(a_{1,0}(0)(\Omega+\frac{\delta}{\hbar} w^2)+2\Omega w^2b_{1,0}(0))+a_{1,0}(0)g_d\cos(\frac{\Omega d t}{2w^2}) \big)\,,\\
	a_{0,1}(t)&=\frac{1}{g_d}e^{-\frac{it}{2w^2}(\Omega(1+m_r)+2(E_{ph}+E_{ex}) w^2)}\big(-i\sin(\frac{\Omega d t}{2w^2})(a_{0,1}(0)(g+\frac{\delta}{\hbar} w^2)+2\Omega w^2b_{0,1}(0))+a_{0,1}(0)g_d\cos(\frac{\Omega d t}{2w^2}) \big)\,,\\
	b_{0,0}(t)&=\frac{1}{g_c}e^{-\frac{it}{4w^2}(\Omega(1+m_r)+2(E_{ph}+E_{ex}) w^2)}\big( i(b_{0,0}(0)\Omega+2w^2(b_{0,0}(0)\frac{\delta}{\hbar}-2a_{0,0}(0)g))\sin(\frac{\Omega c t}{4w^2})+b_{0,0}(0)g_c\cos(\frac{\Omega c t}{4w^2})\big)\,,\\
	b_{1,0}(t)&=\frac{1}{g_d}e^{-\frac{it}{2w^2}(\Omega(1+m_r)+2(E_{ph}+E_{ex}) w^2)}\big(i\sin(\frac{\Omega d t}{2w^2})(b_{1,0}(0)(\Omega+\frac{\delta}{\hbar} w^2)-2\Omega w^2a_{1,0}(0))+b_{1,0}(0)g_d\cos(\frac{\Omega d t}{2w^2}) \big)\,,\\
	b_{0,1}(t)&=\frac{1}{g_d}e^{-\frac{it}{2w^2}(\Omega(1+m_r)+2(E_{ph}+E_{ex}) w^2)}\big(i\sin(\frac{\Omega d t}{2w^2})(b_{0,1}(0)(\Omega+\frac{\delta}{\hbar} w^2)-2\Omega w^2a_{0,1}(0))+b_{0,1}(0)g_d\cos(\frac{\Omega d t}{2w^2}) \big)\,,
	\end{align}
\end{subequations}
\end{widetext}
  
where $\delta=E_{ph}-E_{ex}$ and we introduce:
\begin{subequations}
	\begin{align}
	g_c\equiv&\sqrt{g^2+4\Omega w^2\frac{\delta}{\hbar}+4w^4(4\Omega^2+(\frac{\delta}{\hbar})^2)}\,,\\
	g_d\equiv&\sqrt{\Omega^2+2\Omega w^2\frac{\delta}{\hbar}+w^4(4\Omega^2+(\frac{\delta}{\hbar})^2)}\,,
	\end{align}
\end{subequations}

We present in Fig.~\ref{fig:rytuwasateitrgywey89382hjikj} some
examples for variations of $a_{n,m}$ and $b_{n,m}$ coefficients for
two values of $w=30\xi$ and $w=\xi$. One notes that for $w=30\xi$, we
can safely keep $a_{0,0}$, $a_{0,1}$, and $a_{1,0}$, as they are four
orders larger than the other coefficients. However, with $w=\xi$, we need
to take into account more terms for the approximation of the solution.
\bibliography{dc}
\end{document}